# A categorization of arguments for counting methods for publication and citation indicators[1]

Marianne Gauffriau[a, b]

[a] Faculty Library of Natural and Health Sciences, Copenhagen University Library, The Royal Library, Copenhagen, Denmark, [b] SUND Research & Innovation, Faculty of Health and Medical Sciences, University of Copenhagen, Copenhagen, Denmark, e-mail: mgau@kb.dk, marianne.gauffriau@sund.ku.dk

## Abstract

Most publication and citation indicators are based on datasets with multi-authored publications and thus a change in counting method will often change the value of an indicator. Therefore it is important to know why a specific counting method has been applied. I have identified arguments for counting methods in a sample of 32 bibliometric studies published in 2016 and compared the result with discussions of arguments for counting methods in three older studies. Based on the underlying logics of the arguments I have arranged the arguments in four groups. Group 1 focuses on arguments related to what an indicator measures, Group 2 on the additivity of a counting method, Group 3 on pragmatic reasons for the choice of counting method, and Group 4 on an indicator's influence on the research community or how it is perceived by researchers. This categorization can be used to describe and discuss how bibliometric studies with publication and citation indicators argue for counting methods.

## Keywords

## 1 Introduction

Publication and citation indicators are often used in policy reports about research and many bibliometric research studies focus on the development of new indicators. The choice of counting method is an inevitable and important step in calculating an indicator. A wide range of counting methods can be used to allocate credit for a publication and its citations to the authors, to the authors' institutions, to the journals they have published in, etc. It is well-documented that for multi-authored publications, a change in counting method will often change the value of an indicator and sometimes the conclusion of an analysis (see e.g. Gauffriau *et. al.*, 2008). Therefore, to fully understand an indicator, it is important to know why a specific counting method has been applied. In this paper, I will show that there are at least four groups of arguments for counting methods.

My objective is twofold:

---

[1] A preliminary version of this study was presented at a poster session at the *21st International Conference on Science and Technology Indicators*. València, Spain, 2016.

1. to identify arguments for counting methods for publication and citation indicators in a sample of recent bibliometric studies
2. to use the identified arguments to develop a categorization of arguments for counting methods for publication and citation indicators.

I aim to facilitate the discussion concerning the application of counting methods in bibliometric studies and policy reports. The categorization of arguments for counting methods can help researchers describe and discuss their choice of counting method in a specific study. This information can be used by other researchers and policymakers when assessing and using the study. The categorization can also offer an overview of how counting methods are used across bibliometric studies. Ultimately, how we argue for counting methods may reveal tacit knowledge about what publication and citation indicators measure and add to a common understanding of these indicators.

There is a large body of literature discussing counting methods for publication and citation indicators with a focus on theoretical and methodological arguments for each counting method. This is elaborated in Section 2. To the best of my knowledge, however, there has been no systematic exploration of how researchers choose counting methods for their studies, that is, which arguments they use for a counting method. This is what I wish to analyze to establish a new way of discussing counting methods. The aim of my study is not to provide a literature review on counting methods, which would probably mainly encompass theoretical and methodological arguments. Nor is it to identify the most appropriate counting method as this would probably draw upon just one or two types of arguments and a specific purpose of an analysis. As my analysis will show, there are many very different types of arguments for counting methods, and these can be grouped in accordance with their distinct underlying logics.

## 2 Theory

### 2.1 Terminology

Throughout the paper, I use the terminology established by Gauffriau *et al.* (2007, pp. 178-180), except that I replace the term normalized with fractionalized. Following this terminology, whole counting means that all authors of a publication get one credit each for the publication. If countries are the basic unit of analysis, all unique countries mentioned in the affiliation of a publication get one credit each. This counting method is called full/total/integer/whole counting, or simply number of publications/citations, or counts in many of studies in my sample. Complete fractionalized counting means that all authors of a publication share one credit for the publication. If countries are the basic unit of analysis, all countries mentioned in the affiliation of a publication share one credit. In the sample of studies, this method is often called fractional counting. Complete fractionalized counting can be divided in two sub-groups: the rank-independent in which all authors of a publication get equal shares of the credit, and the rank-dependent in which each author gets a share of the credit depending on his/her position in the byline. An example of the latter is straight counting, in which the first author gets one credit and co-authors get none. There are many other types of complete fractionalized and rank-dependent counting methods, e.g. harmonic counting, arithmetic counting, geometric counting (see a short review in Waltman, 2016, p. 379-380).

## 2.2 Background

Counting methods were discussed even in the pioneering scientometric works. In the definition of Lotka's law, Lotka wrote about the counting method: "Joint contributions have in all cases been credited to the senior author only" (Lotka, 1926, p. 323) and advanced the idea that some publications "should perhaps be considered separately since they are not the product of one person unassisted"(ibid.). Price analyzed whole, complete fractionalized, and straight counting in *Big Science, Little Science* (Price, 1963, pp. 127-129), and Cole and Cole discussed whole and straight counting in *Social Stratification in Science* (Cole & Cole, 1973, pp. 32-33).

Today, there is still a debate concerning counting methods. Counting methods are compared, and the best is selected, or new counting methods are developed. In contrast to these approaches, I focus on the arguments for the counting methods. Studies that discuss counting methods argue for counting methods too, of course, but it is the counting methods themselves that are analyzed, not the arguments. Below, I give a few examples of arguments for counting methods from one of the first studies to analyze the effect of different counting methods.

Lindsey analyzed and discussed straight, whole, and complete fractionalized counting and ultimately advocated complete fractionalized counting for publication and citation indicators as a consequence of the growing number of multi-authored publications (Lindsey, 1980). Many arguments can be identified in the analysis and discussion. Lindsey did not agree with all of them. A pragmatic argument for straight counting is that the method "greatly reduces the work required to collect data" (ibid., p. 146). An argument for complete fractionalized counting refers to mathematical properties: "the weights must sum to one, because what is being measured is 'one scientific paper'" (ibid., p. 151). The next sentence in Lindsey's study is an argument for whole counting based on intuition: "Although this [fractionalization] is the logical procedure, it violates the intuitive judgment of many scientists" (ibid., p. 151). The examples show that the intention behind the choice of counting method can point in many directions. If this information is not available in a bibliometric study, it can lead to misinterpretation of an indicator.

A few studies have presented a dedicated analysis of and discussion on how bibliometric studies argue for counting methods. Larsen concluded that only five out of 85 studies from the *International Society for Scientometrics and Informetrics'* 2005 and 2007 proceedings argued for the choices of counting methods (Larsen, 2008, p. 237). Gauffriau *et al.* presented examples of arguments from the literature and discussed them from a mathematical perspective (Gauffriau *et al.*, 2008, pp. 161-169). Waltman and van Eck reported their discussions with bibliometricians (researchers and practitioners) on arguments for and against whole and complete fractionalized counting: Which of the two methods is the more intuitive? Or do the two counting methods measure participation and contribution respectively? Or how is contribution and collaboration measured? (Waltman & van Eck, 2015, pp. 889-890). In contrast to these studies, I do not seek to argue for a specific counting method as the right choice for a specific bibliometric analysis but simply to identify categories of arguments for counting methods.

# 3 Material and methods

## 3.1 Data

I have analyzed a sample of 32 bibliometric studies published in 2016 that meet two criteria: 1) having publication and citation indicators as part of the method and result sections, and 2) arguing explicitly for the choices of counting methods. I do not interpret an explicit argument against a counting method as an implicit argument for the opposite counting method because this would require a system of dichotomies. For example, what is the opposite of straight counting?

The sample is from peer reviewed journals and is, as such, approved by peers in the research field. Furthermore, journal publications should give enough information for other researchers to evaluate a study. I thus expected to find discussions of methods and results and potentially arguments for counting methods.

Table 1
Journals searched to identify studies for inclusion in the analysis

|  | First issues from 2016, first issues from the second half of 2016, and last issues from 2016: | |
| --- | --- | --- |
|  | *Journal of Informetrics,* and *Scientometrics* | *Research Policy, Research Evaluation, Journal of the American Society for Information Science and Technology (JASIST), Aslib Journal of Information Management, Information Processing and Management, Journal of Documentation,* and *Journal of Information Science* |
| All studies | 166 studies | Not analyzed |
| Studies that meet criteria 1) | 99 studies | Not analyzed |
| Studies that meet criteria 1) and criteria 2) | 26 studies | 6 studies |
| All analyzed studies | 32 studies | |

The list of journals was taken from a literature review on citation impact indicators (Waltman, 2016, p. 367).

Twenty-six studies in the sample are from *Journal of Informetrics* and *Scientometrics* (Table 1). The respective scopes of the two journals are "topics in bibliometrics, scientometrics, webometrics, and altmetrics"[2] and "investigations in which the development and mechanism of science are studied by statistical mathematical methods"[3]. I selected the studies in accordance with the two aforementioned criteria. If other types of indicators were present in a study, I only analyzed the part about publication and citation indicators. Other studies (e.g. network analysis, altmetrics, and patent analysis) were excluded from the analysis. Furthermore, a publication can be assigned to multiple subject areas. I did not include arguments for counting methods in relation to this. Finally, I only analyzed arguments for individual counting methods and not for the combination of counting methods. For example, the difference between complete fractionalized and whole scores can be used as an indicator for collaboration. For many of the excluded studies, it would indeed be relevant to discuss arguments for counting methods, but this argumentation might follow a different logic than arguments for counting methods for publication or citation indicators included in my study.

---

[2] http://www.journals.elsevier.com/journal-of-informetrics/ [Accessed: September 26th, 2016]
[3] http://link.springer.com/journal/11192 [Accessed: September 26th, 2016]

As a supplement, I searched the following journals and applied the same two criteria as mentioned above: *Research Policy, Research Evaluation, Journal of the American Society for Information Science and Technology (JASIST), Aslib Journal of Information Management, Information Processing and Management* (Issues 1 and 2, 2016, as Issue 1 was a special issue on social media)*, Journal of Documentation,* and *Journal of Information Science* (Issues 1 and 2, 2016, as Issue 1 was a special issue on search as learning). As shown in Table 1, this resulted in the inclusion of six additional studies in the analysis.

### 3.2 Identification and categorization of arguments

The selection of studies from *Journal of Informetrics* and *Scientometrics* was done as follows. In total 166 studies from the six 2016 issues were saved in NVivo 11 Pro, a commercial program for qualitative data analysis (http://www.qsrinternational.com/nvivo-product/nvivo11-for-windows/pro). I skimmed the 166 studies and did a preliminary selection of which studies to include in the analysis. Relevant text phrases were coded under the labels: indicator, counting method, arguments for counting method, object of study, and database. From the coded text in the first 2016 issues of the journals, I identified keywords (Appendix A) and the keywords were saved as highlighted text in the 166 studies. I went through the 166 studies again and, based on the highlighted text, identified more text phrases for coding. Of the 166 studies, 99 met the criteria of having publication and citation indicators as part of the method and result sections. Of these, 26 studies met the criteria of having explicit arguments for the choices of counting methods (Table 1 and Appendix B).

For journals other than *Journal of Informetrics* and *Scientometrics*, I skimmed titles and abstracts and found studies with publications and citation indicators. This technique was used as these journals have a much broader scope than do *Journal of Informetrics* and *Scientometrics*, and only few of the studies were relevant for my purpose. I skimmed the full texts of studies with publication and citation indicators and identified the six studies that met both criteria for inclusion (Table 1 and Appendix B). These were also searched for keywords (Appendix A) and coded.

In total 32 studies met both criteria for inclusion in the analysis. I identified arguments for counting methods and assigned the arguments to categories. Some authors only argued for the counting method they used while others discussed multiple counting methods and ended up choosing the one(s) they found most appropriate. In my analysis, I have included all arguments and not only the arguments for the counting methods actually used in the studies. A study can thus be represented in several categories and argue for several counting methods.

The categorization of arguments for counting methods was done based on quotations from the 32 studies and no further guidelines for what a counting method measures or how it should be used. To make the analysis as transparent as possible I show all quotations in Tables 2.1.1-2.4. Finally, I suggest how the categories can be divided in four groups in accordance with the underlying logics of the arguments (Table 3).

The categorization of arguments for counting methods for publication and citation indicators represents a snapshot of the state of art based on a sample of bibliometric studies published in 2016. To validate the categorization, I compared it with the arguments discussed in older studies: Gauffriau *et al.* (2008, pp. 161-169), Larsen (2008), and Waltman and van Eck (2015, pp. 889-891).

### 3.3 Limitations

The qualitative text analysis probably favors some categories of arguments over others. My analysis indicates that many bibliometric studies use whole counting without any justification and furthermore that pragmatic arguments are often linked to whole counting. Thus, pragmatic arguments are represented in the analysis, but they may be underrepresented.

My focus is to develop categories of arguments and assign counting methods to the categories. A more fine-grained analysis could be carried out, but this is beyond the scope of my analysis as it would require a huge sample of studies or a focus on one or a few variables to get a manageable sample. To refine the analysis, the object of study could be added as a variable: Does the argument for a counting method change if the object of study changes from country to author? This is not analyzed in my study. Another possibility is to look at publication and citation indicators separately. I assume that an argument for a counting method for publication indicator is also valid for a citation indicator and vice versa. Furthermore, the study is limited to publication and citation indicators. Many other indicators and methods also depend on counting methods, and a similar categorization of arguments for counting methods could be developed for those, for example, for publications with multiple subject areas, altmetric indicators, and network analysis. My expectation is that some categories will prove valid across different indicators and methods while others will not.

## 4 Results and discussion

### 4.1 Overview of analyzed studies

I start with an overview of the 99 studies from *Journal of Informetrics* and *Scientometrics*, that meet criteria 1) having publication and citation indicators as part of the method and result sections. For 32 studies, the choice of counting method is immaterial as there is only one possible value for an object of study (e.g. one journal title or one author) per publication (see Appendix B). Therefore, a change of counting method has no effect.

For the remaining 67 studies, a change of counting methods will affect the value of the publication and citation indicators. In each study, I have identified the counting method applied. The studies describe the counting methods in words, through a formula, give examples of the calculations, give references to well-defined indicators from specific databases, and/or give references to studies in which the method is defined. Of the 67 studies, 48 use whole counting; four use complete fractionalized counting; and 15 use whole, complete fractionalized, straight counting, etc. in different combinations.

Twenty-six studies explicitly justify the choice of counting method and thus provide arguments for one or more counting methods (criteria 2). Of these, 11 studies use whole counting; four[4] use complete fractionalized counting; and 11 use whole, complete fractionalized, straight counting etc. in different combinations.

### 4.2 Categorization of arguments for counting methods

The Tables 2.1.1-2.4 show the arguments from the 26 studies from *Journal of Informetrics* and *Scientometrics* plus the six studies from other journals. I have assigned the arguments to categories. There is one table per category. The categories are grouped: Tables 2.1.1-2.1.5, Table 2.2, Tables 2.3.1-2.3.4, and Table 2.4.

As mentioned in Section 3, I do not distinguish between counting methods for publication versus citation indicators. Twenty-one of the 32 studies analyzed include citation indicators, and overall these indicators are described as measuring impact.

For a few of the arguments, I made an interpretation of the text guided by the intentions of the studies. My interpretations are reported in the tables.

Table 2.1.1
Category: The indicator measures (the impact of) participation

| **Arguments for whole counting** |
|---|
| "…full counting method can thus be seen as measuring participation…" (Cimini *et al.*, 2016, p.204) |
| "Since the participation in a research area is determined by whether a core paper (or a top 10 % citing paper) of a target country is included in a research area, we used the number of publications counted by the whole counting…" (Igami & Saka, 2016, p. 392) |
| "Here we show that the quantification of double-counting articles can indirectly indicate the proportion of country's participation in the publications of a particular field." (Zanotto *et al.,* 2016, p. 1796) |

Table 2.1.2
Category: The indicator measures (the impact of) production

| **Argument for whole counting** |
|---|
| "We have undertaken the whole method for the calculation of productivity of authors." (Kumar, 2016, p. 22) |
| "Number of publications […] Aim to Assess […] Production" (Wildgaard, 2016, p. 1073) |

---

[4] The counting methods and arguments are the same in the three studies and therefore only reported once in Table 2.1.3. The four studies are: Abramo, G. et al., (2016) *Journal of Informetrics*, Volume 10, Issue 1, pp. 31-42; Abramo, G. et al., (2016) *Journal of Informetrics*, Volume 10, Issue 3, pp. 854-862; Abramo, G. et al., (2016) *Scientometrics*, Volume 106, Issue 1, pp. 119-141; and Abramo, G. et al., (2016) *Scientometrics*, Volume 109, Issue 3, pp. 2093-2117.

| |
|---|
| "…large differences in productivity across fields, depending on how productivity is defined. The 'hard sciences' (natural, technical, and medical sciences) produce a much large number of publications per person and have a higher number of coauthors. However, in terms of fractional counts, the 'soft sciences' (social sciences and humanities) are more productive." (Bloch & Schneider, 2016, p. 378) |
| **Arguments for complete fractionalized counting** |
| "We use fractional counted productivity […] to account in some imprecise way for the number of coauthors." (van den Besselaar & Sandström, 2016, p. 157) |
| "Fractional publications […] Aim to Assess […] Production if the author had worked alone" (Wildgaard, 2016, p. 1073) |
| "Then we compute the number of fractional publications to the number of professor in each group in order to find the fractional productivities." (Yuret, 2016, p. 1199) |
| "…large differences in productivity across fields, depending on how productivity is defined. The 'hard sciences' (natural, technical, and medical sciences) produce a much large number of publications per person and have a higher number of coauthors. However, in terms of fractional counts, the 'soft sciences' (social sciences and humanities) are more productive." (Bloch & Schneider, 2016, p. 378) |
| "While it would appear important to take account of the number of coauthors (fractional counts), measuring individual researcher productivity is a still a complex task (see e.g. Abramo, Cicero and D'Angelo 2013; Abramo, D'Angelo and Rosati 2013)." (Bloch & Schneider, 2016, p. 379) |
| "Chemistry articles typically have more authors than economics articles. Therefore, the number of authors is one of the causes that creates the differences in productivities between economics and chemistry. In order to account for this, we compute the fractional publications. That is, we attribute 1/n publications to each author in an n author paper." (Yuret, 2016, p. 1199) |

Table 2.1.3
Category: The indicator measures (the impact of) contribution

| |
|---|
| **Arguments for whole counting** |
| "…reasonable to accept the simplification that all authors' contributions are equally important. […] the different types of collaboration should be considered as possible explanations for any patterns found." (Thelwall & Sud, 2016, p.49)<br><br>This study can also be assigned to Table 2.3.3. |
| "Although fractional counting may more accurately indicate the contribution by each author, this issue is often debated. For example, in certain papers the first author may have done the major portion of the work. In other instances, it may happen that all co-authors had contributed equally. However, these aspects are rarely disclosed, hence bibliometric data are unable to trap finer nuances. For this reason, the whole paper counting method is often preferred" (Kumar, 2016, pp. 22-23) |
| **Arguments for complete fractionalized counting** |
| "…fractional contribution of researcher to publication…" (Abramo *et al.*, 2016, p. 34) |
| "Fractional authorship quantifies an individual author's contributions to published papers, and it can be used to estimate the number of papers written by a particular scientist." (Pritychenko, 2016, p. |
| 464) |
| "…the fractions from each group on each paper were calculated. This enabled us to determine the fractional counts of contributions from each of the five groups to any given set of papers, such as those from a triennium, from a state or federal territory, or from a major field." (Lewison *et al.,* 2016, 1883-1884) |
| "If, in author-level fractionalization, a publication with e.g. three co-authors corresponds to one-third of a single-authored publication, the implicit assumption is made that all publications in a field reflect the same amount of effort, originality, and so on, and that the number of authors directly corresponds to the amount of their respective contributions. This is a plausible, but not actually evidence-based premise. […] Fractionalization on the level of author addresses can be seen as a proxy for the respective author contributions…" (Möller *et al.,* 2016, p. 2222) |
| **Arguments for complete fractionalized counting, rank-independent** |
| "…authors in simple alphabetical order: in this case the fractional contribution simply equals the inverse of the number of authors." (Abramo *et al.*, 2016, p. 34) |
| "…researchers' individual publication output was […] operationalized as the individual sum of publication shares. […] A single publication share is also called 'contribution' or 'publication-equivalent' and is based on the assumption that 'that each person contributed the same amount' to a publication…" (Fell & König, 2016, p. 121) |
| **Arguments for complete fractionalized counting, rank-dependent** |
| "…indicate the contributions to the published research by the order of the names in the listing of the authors. […] giving different weights to each co-author according to their position in the list of authors and the character of the co-authorship (intra-mural or extra-mural)" (Abramo *et al.*, 2016, p. 34) |
| "The arithmetic fraction is based on an equal distribution of authorship credit among coauthors, while in nuclear physics, it is reasonable to assume that the first author usually contributes ≈ 50 %, and the rest is split between the other coauthors." (Pritychenko, 2016, p. 465) |

Table 2.1.4

Category: The indicator measures (the impact of) output/volume/creditable to/performance

| |
|---|
| **Arguments for whole counting** |
| "…researchers' individual publication output was operationalized as the absolute number of publications that were (co-)authored by them…" (Fell & König, 2016, p. 121) |
| **Arguments for complete fractionalized counting** |
| "…indicator is the shares (based on the fractional counting) of Japan and benchmarking countries in core papers from all research areas… […] This is an indicator that shows the volume of papers published in hot research areas." (Igami & Saka, 2016, p. 389) |
| "…fractional counting as measuring how many papers are creditable to a country…" (Cimini *et al.*, 2016, p.204) |
| "…if a publication is coauthored by two or more scientists, the evaluator may choose to count that publication as half or 1/(number of coauthors), when ranking the research performance of one of the coauthors." (Frittelli *et al.,* 2016, p. 3062) |

Table 2.1.5

Category: The indicator measures (the impact of) the role of authors

| **Arguments for straight counting** |
|---|
| "Further analysis of the 93,867 nuclear physics authors shows only 43,037 individuals who have published first-author papers, including 35,338 non-alphabetical author lists. These results highlight two distinct groups of authors: project leaders (risk takers) and followers." (Pritychenko, 2016, p. 466) |
| "…the index value was calculated based on first affiliations listed for each article, i.e. countries of the lead authors." (Klincewicz, 2016, p. 335) |
| "The number of first-authored articles in foreign-language, refereed journals presumably captures an original academic contribution as a main contributor, rather than as a manager of a laboratory." (Kawaguchi *et al.*, 2016, p. 1440) |
| **Argument for last author counting** |
| "…one finds the person in charge in the last position. It stands to reason that […] group leaders or principal investigators should be found in the last position of the list of authors" (Neufeld, 2016, p. 57) |
| **Argument for reprint author counting** |
| "Most citing articles have multiple authors, potentially located in different countries. We use the reprint author to determine the location of the citing article because we assume that this author is more likely to be closely connected to the research than a randomly chosen author. This assumption is based on our experience reading the bibliographic information for publications on Web of Science, in which the reprint author commonly was the first or last author." (Kahn & MacGarvie, 2016, p. 1306) |

Table 2.2

Category: Additivity of counting method

| **Arguments for whole counting** |
|---|
| "…the non-additive character of scientific publications should be recognized. Direct comparisons or summing up publications with different authors, organizations, or journals is problematic… […] the present study was very conservative in using the data, i.e. no aggregate constructs were derived from the bibliometric data…" (Klincewicz, 2016, p. 342) |
| "Double occurrences were excluded within each unit of analysis: a paper co-authored by two or more researchers belonging to the same institution was counted only once, and one paper authored by two or more institutions was counted once at a higher aggregation level (Portugal)." (Ramos & Sarrico, 2016, p. 97) |
| "Using whole counts, e.g. if one of the addresses is in country C, this country receives one score […] We also calculated the h-core score of the United Kingdom (UK), taken care not to double count collaborations between England, Northern Ireland, Scotland and Wales." (Sanz-Casado et al., 2016, p. 257) |
| "For this analysis we eliminate the double counts of publications co-authored by professors |

pertaining to the same GEV [panel of experts]. We consider instead publications co-authored by professors of different GEVs, because each GEV adopts different thresholds." (Abramo & D'Angelo, 2016, p. 2059)

| **Arguments for complete fractionalized counting** |
|---|
| "…works (Aksnes et al., 2012; Waltman and van Eck, 2015) have argued that fractional counting […] leads to a proper field normalization of impact indicators…" (Cimini *et al.*, 2016, p. 204) <br><br> I have assigned the argument to Table 2.2 as the references in the quotation conclude that complete fractional counting of authors/affiliations is additive. |
| "…fractionate these people's contributions by country in order to make the sum of the individual contributions equal the number of papers…" (Lewison *et al.*, 2016, p. 115) |
| "For reasons of mathematical consistency, the fractionalized approach is preferable, as Waltman and van Eck (2015) demonstrate, and Aksnes et al. (2012) also point out." (Möller *et al.*, 2016, p. 2222) <br><br> I have assigned the argument to Table 2.2 as the references in the quotation conclude that complete fractional counting of authors/affiliations is additive. |
| "Traditionally, when calculating a (field normalized) citation score using fractional counting, each paper is given an aggregate weight of one (1), which weight is split between authors or organisations depending on level of fractionalisation. Hence, when calculating the (field normalized) citation score of a university, the university's papers in social science has been given the same weight per paper as the university's papers in chemistry." (Koski *et al.*, 2016, p. 1151) |

Table 2.3.1

Category: Availability of data

| **Argument for whole counting** |
|---|
| "…full counting is adopted as the SCImago statistics are built according to this criterion." (Cimini *et al.*, 2016, p. 204) |
| **Argument for straight or reprint author counting** |
| "The restriction to first authors was necessary since other authors, in the case of collaborative papers, could not be associated with specific addresses. […] Complete author-address information only exists in Web of Science after 2008. An alternative solution is to use reprint authors (which reveal similar results)." (Koski *et al.*, 2016, p. 1148) |

Table 2.3.2

Category: Prevalence of counting method

| **Argument for whole counting** |
|---|
| "…full counting is also commonly adopted…" (Cimini *et al.*, 2016, p. 204) |
| "On the level of international comparison, the whole count method is generally used, but fractionalized approaches are increasingly being advocated." (Möller *et al.*, 2016, p. 2222) |

Table 2.3.3

Category: Simplification of indicator

| **Arguments for whole counting** |
|---|
| "…the data contains no overlap of individual networks of different reference institutions or at least it was possible to ignore this overlap; […] Multiple co-authorships with authors from different institutions create duplicates of papers [overlap] […] Measurement dependencies [overlap] are taken into account to some extent by a multilevel statistical modelling strategy." (Bornmann *et al.*, 2016, p. 316) |
| "For simplicity only the number of papers and citations is captured […] In other words, the total number of papers and citations received are the parameter to be evaluated …" (Höylä *et al.*, 2016, p. 266) |
| "For simplicity, we used the normal count of articles with at least one author affiliated to Japanese universities." (Morichika & Shibayama, 2016, p. 225) |
| "…for simplicity reasons we use the raw numbers…" (Piro & Sivertsen, 2016, p. 2269) |
| "In this work, counted articles for each country includes the publications resulting from international collaboration. Unfortunately due to the large numbers of articles and countries in this study, fractional counting as recommended by Gaufrial and Larsen 2005, could not be applied here." (Zanotto *et al.,* 2016, p. 1794) |

Table 2.3.4

Category: Insensitive to change of counting method

| **Argument for whole counting** |
|---|
| "In this work, full counting is adopted […] …the difference between the two methods [whole and complete fractionalized] basically consists in a country-specific rescaling of impact indicators, the relative temporal changes of countries impact we will analyse are, per se, unaffected by the choice of the counting method." (Cimini *et al.*, 2016, p. 204) |

Table 2.4

Category: Incentive against collaboration

| **Argument for complete fractional counting** |
|---|
| "The effect of fractionalization is that collaborations are virtually disadvantaged—which is especially precarious when it comes to measuring intersectorial collaborations… […] On the other hand, however, it seems reasonable to fractionalize in order to neutralize an increasing international tendency toward collaboration." (Möller *et al.,* 2016, p. 2222) |

A few of the 99 studies discussed how publications with varying numbers of co-authors may affect the value of a publication or citation indicator, see e.g. Valentin *et al.* (2016, p. 77). These studies are not included in the tables above as I found no arguments for the choices of counting methods. Counting methods were not reported as relevant for the discussion on multi-authored publications. Along similar lines, one could try to uncover the arguments for the choices of counting methods in the studies that did not meet criteria 2). This would be based on the indicator, the counting method, what the study proposes to measure, the database used, etc. This was not done as the method did not seem convincingly systematic and reliable. In other words, there was insufficient information in the studies to carry out this exercise.

To validate the categorization, I compare the Tables 2.1.1-2.4 with the arguments discussed in Gauffriau *et al.* (2008, pp. 161-169), Larsen (2008), and Waltman and van Eck (2015, pp. 889-891). Most of the categories from the tables are also present in these three studies. This indicates that frequently used arguments are covered. But the overlap is not complete and this may indicate that not all possible arguments for all counting methods are identified. Because new indicators are invented and new data sources are made available all the time, it is impossible to develop a final set of categories.

Table 2.4 includes one argument and is not grouped with other categories. However, when I compare with arguments in Gauffriau *et al.* (2008) and Waltman and van Eck (2015), two categories are discussed that have a similar underlying logic. One is: "Fractional counting provides an incentive against collaboration" (Waltman & van Eck, 2015, p. 889). The other is: "The distinction between actual [complete fractionalized] and perceived [whole] publications and citations was introduced with the argument that the critical issue is how the scientific community perceives a publication with respect to each of its authors" (Gauffriau *et al.*, 2008, p. 167) and "full counting is in agreement with the intuitive idea that all publications of a researcher or a research group should be considered of equal importance" (Waltman & van Eck, 2015, p. 891). These categories are included in the analysis together with Table 2.4.

Table 3 presents an overview of the identified categories and the counting methods for which they argue. I have furthermore grouped the categories in accordance with their underlying logics.

Table 3
Categorization of arguments for counting methods for publication and citation indicators

| Category | Counting method(s) |
|---|---|
| **Group 1: The indicator measures (the impact of) …** | |
| … participation of an object of study (Table 2.1.1) | Whole |
| … production of an object of study (Table 2.1.2) | Whole, complete fractionalized |
| … contribution of an object of study (Table 2.1.3) | Whole, complete fractionalized (rank-independent and rank-dependent) |
| … output/volume/creditable to/performance of an object of study (Table 2.1.4) | Whole, complete fractionalized |
| … the role of authors affiliated with an object of study (Table 2.1.5) | Straight, last author, reprint author |
| **Group 2: Additivity of counting method** | |
| Additivity of counting method (Table 2.2) | Whole, complete fractionalized |
| **Group 3: Pragmatic reasons** | |
| Availability of data (Table 2.3.1) | Whole, straight, reprint author |

| Prevalence of counting method (Table 2.3.2) | Whole |
|---|---|
| Simplification of indicator (Table 2.3.3) | Whole |
| Insensitive to change of counting method (Table 2.3.4) | Whole |
| **Group 4: Influence on/from the research community** | |
| Incentive against collaboration (Table 2.4) | Complete fractionalized |
| Comply with researchers' perception of how their publications and/or citations are counted | Whole |

## 4.3 Discussion of the four groups with arguments for counting methods

I will now discuss the underlying logics of the four groups and the counting methods linked to the groups. Please note that the discussion is based on a limited number of studies and does not cover all counting methods or all possible arguments for counting methods.

For Group 1, the underlying logic for the argumentation for a counting method relates to the concept that the study attempts to measure. In Group 1 the concepts participation (Table 2.1.1), production (Table 2.1.2), contribution (Table 2.1.3), and role of authors (Table 2.1.5) have separate tables as these concepts were mentioned in multiple studies included in my analysis. Other concepts such as volume, output, etc. were only mentioned once and are reported in Table 2.1.4. For the category with most arguments (Table 2.1.3), there seems to be no consensus on the appropriate counting method. This may indicate that the concept is poorly operationalized across bibliometric studies. What do we measure when we measure contribution? Even though I cannot answer this question based on my analysis, it is evident from Group 1 that counting methods are used to operationalize concepts in bibliometric studies.

For Group 2 (Table 2.2) the underlying logic relates to the mathematical properties of the counting method itself: namely, to ensure that the counting method is additive and to avoid double counting of publications and citations. There seems to be consensus among the analyzed studies that complete fractionalized counting is additive, and whole counting is non-additive, which can lead to double counting if scores for more objects of study are added.

Group 1 and 2 can be seen as building upon theoretical/methodological arguments.

Group 3 includes four categories (Tables 2.3.1-2.3.4). These do not take into account the theoretical/methodological arguments discussed above but are linked to pragmatic reasons. The categories in Group 3 are except for one argument linked to whole counting. This can be explained by the fact that whole counting is the readily available counting method in databases often used to calculate bibliometric indicators (Web of Science, Scopus, and Google Scholar). For example, a search for publications from Denmark will return the number of publications in which Denmark is present at least once in the list of affiliations. This corresponds to whole counting. The category on availability of data (Table 2.3.1) is thus assigned to Group 3. In Table 2.3.1, one argument is for straight/reprint author counting but follows the same underlying logics as the arguments for whole counting. In continuation of this, whole counting is often used for publication and citation indicators. I showed that 48 of 67 studies in *Journal of Informetrics* and *Scientometrics* relied on whole counting alone. The category on prevalence of a counting method (Table 2.3.2) is therefore also assigned to the group. Furthermore, this may be the reason why arguments for whole counting refer to the category on simplification of an indicator (Table 2.3.3)*.* This is how I interpret most of the quotations in Table 2.3.3. Whole counting is easier to work with and to understand. I regard Bornmann *et al.*'s argument for Simplification as slightly different as they want to limit the number of variables and focus on other properties of an indicator. They discuss what this means for the analysis. Finally, the category on study conclusions that are insensitive to a change of counting method (Table 2.3.4) is included in the group. It is used in a study in which the value of an indicator will change if the counting method is changed, without, however, affecting the conclusion of the study. It is easier to use whole counting as explained above, with the result that the category can be assigned to Group 3.

For Group 4 (Table 2.4 and categories from Gauffriau *et al.*, 2008, and Waltman & van Eck, 2015.) the underlying logic is not related to what an indicator measures (Group 1) but instead to the impact of the indicator on the research community under evaluation and vice versa. Two categories are included in the group. One is to Comply with researchers' perception of how their publications and/or citations are counted and is sometimes mentioned as being of particular importance if the object of study is individual researchers (Waltman & van Eck, 2015, p. 891). The category is linked to whole counting and corresponds to counting the publications in a researcher's publication list. The focus of the other category is to affect the behavior of the research community by, for example, introducing an incentive against collaboration. In Table 3, the category is linked to complete fractionalized counting. If other incentives are pushed, other counting methods may very well be chosen.

One could say that a sound bibliometric analysis should avoid arguments in Group 3, Pragmatic reasons. But if we wish to discuss how counting methods are used to calculate publication and citation indicators, all groups are present and must be taken into account. In *Journal of Informetrics* and *Scientometrics*, 26 of 67 studies explicitly argued for the choices of counting methods. For the remaining 41 studies, the counting methods could be determined, but it would be of great value also to know why a counting method was chosen. If other researchers or policymakers wish to use one of the 41 studies it will be important to know the intention behind the choice of counting method. For example, was whole counting chosen because it was easily available or because the indicator was carefully designed to measure participation of an object of study?

A pragmatic argumentation for a counting method can be fully justified. For example, in studies in which a bibliometric indicator or method is discussed, the counting method can be downplayed if other properties are in focus. Still, it is important to declare which counting method is used and ideally also to discuss the choice as this information can be important for subsequent studies. See, for example, Lewison *et al.* (2016), who developed a method for establishing the gender and nationality of authors on the basis of their names. Whole counting (actually without an explicit argument for the counting method) is used, but as a limitation it is mentioned that "for many purposes it would be more useful to have a fractional count, based on the number of different authors of each paper" (Lewison *et al.*, 2016, p. 115).

# 5 Conclusion

In this paper, I present a categorization of arguments for counting methods for publication and citation indicators. Although the categorization could undergo further elaboration, it provides a new way of discussing counting methods and shows the diversity in arguments for counting methods.

Generally speaking, the categorization of arguments for counting methods can provide new perspectives on what publication and citation indicators measure and how they are understood by considering the (lack of) consensus between the arguments assigned to a category and counting methods supported in the arguments in the category. For example, if studies of the same concept choose different counting methods and no explanation can be found, it may indicate that the concept is poorly operationalized across bibliometric studies.

At the specific level, the categorization of arguments for counting methods can help researchers describe and discuss their choices of counting methods and help users of studies with publication and citation indicators assess and use the studies.

# 6 Acknowledgements

I wish to thank the participants at the poster session at the *21st International Conference on Science and Technology Indicators* for comments on a preliminary version of the results presented in this paper (Table 3). I am also grateful for comments from Peder Olesen Larsen and two anonymous referees. The comments helped me focus the study and improve the analysis.

# 7 Funding sources

This research did not receive any specific grant from funding agencies in the public, commercial, or not-for-profit sectors.

# 8 Appendices

### Appendix A: Lists of keywords

Table A.1
Lists of keywords used to identify studies that meet the criteria of 1) having publication and citation indicators as part of the method and result sections and 2) arguing explicitly for the choices of counting methods.

| Database | Counting method | Basic unit of analysis/ object of study | Other |
|---|---|---|---|
| "Citation Index" | "number of" | academic* | analy* |
| "Google Scholar" | fractional* | affiliat* | assess* |
| "Journal Citation Reports" | full* | area* | assign* |
| "Thomson Reuters" | integer* | article* | attribute* |
| "Web of Knowledge" | multiplicative* | author* | benchmark* |
| "Web of Science" | normali* | categor* | bibliometric* |
| AHCI | raw* | citation* | calculat* |
| CRIS* | total* | cited* | contribut* |
| database* | whole* | classif* | count* |
| Elsevier* | | coauthor* | coverage* |
| JCR* | | co-author* | distribut* |
| Leiden* | | collaborat* | evaluat* |
| repositor* | | disciplin* | frequen* |
| SCI | | document* | impact* |
| SCI-E | | domain* | index* |
| SCImago* | | facult* | indicator* |
| SciVal* | | field* | indices |
| SciVerse* | | international* | measure* |
| Scopus* | | journal* | metric* |
| SIR* | | paper* | monitor* |
| SJR* | | participat* | number* |
| SSCI | | publication* | outcome* |
| WoS* | | publish* | output* |
| | | record* | perform* |
| | | researcher* | position* |
| | | scholar* | product* |
| | | scientist* | receive* |
| | | subject* | score* |
| | | topic* | share* |
| | | | volume* |
| | | | weight* |

**Appendix B: 105 studies that meet the criteria 1) having publication and citation indicators as part of the method and result sections**

**From *Journal of Informetrics*, Volume 10, Issue 1 (February 2016)**
- Abramo, G. *et al.*, pp. 31-42 (Included in the analysis)
- Bonaccorsi, A. *et al.*, pp. 224-237
- Bornmann, L. *et al.*, pp. 312-327 (Included in the analysis)
- Bouyssou, D. *et al.*, pp. 183-199 (A change of counting method has no effect)
- Cimini, G. *et al.*, pp. 200-211 (Included in the analysis)
- Diniz-Filho, J. A. F. *et al.*, pp. 151-161
- Haddawy, P. *et al.*, pp. 162-173 (A change of counting method has no effect)
- Haunschild, R. *et al.*, pp. 62-73
- Letchford, A. *et al.*, pp. 1-8 (A change of counting method has no effect)

- Shen, Z. *et al.*, pp. 82-97
- Thelwall, M., pp. 110-123
- Thelwall, M. *et al.*, pp. 48-61 (Included in the analysis)
- Vieira, E. S. *et al.*, pp. 286-298
- Yang, G. *et al.*, pp. 238-253

**From *Journal of Informetrics*, Volume 10, Issue 3 (August 2016)**

- Abramo, G. *et al.*, pp. 854-862 (Included in the analysis)
- Bornmann, L. *et al.*, pp. 875-887 (A change of counting method has no effect)
- Chi, P.-S., pp. 814-829 (A change of counting method has no effect)
- Malesios, C., pp. 719-731 (A change of counting method has no effect)
- Niu, Q. *et al.*, pp. 842-853
- Saarela, M. *et al.*, pp. 693-718 (A change of counting method has no effect)
- Sidiropoulos, A. *et al.*, pp. 789-813
- Thelwall, M., pp. 863-874 (A change of counting method has no effect)
- Zoller, D. *et al.*, pp. 732-749 (A change of counting method has no effect)

**From *Journal of Informetrics*, Volume 10, Issue 4 (November 2016)**

- Abramo, G. *et al.*, pp. 889-901 (Included in the analysis)
- Colavizza, G. *et al.*, pp. 1037-1051 (A change of counting method has no effect)
- Hu, X. *et al.*, pp. 1079-1091 (A change of counting method has no effect)
- Koski, T. *et al.*, pp. 1143-1152 (Included in the analysis)
- Laakso, M. *et al.*, pp. 919-932 (A change of counting method has no effect)
- Mariani, M.S. *et al.*, pp. 1207-1223 (A change of counting method has no effect)
- Min, C. *et al.*, pp. 1153-1165 (A change of counting method has no effect)
- Onodera, N. *et al.*, pp. 981-1004
- Pooladian, A. *et al.*, pp. 1135-1142 (A change of counting method has no effect)
- Singh, M. *et al.*, pp. 1005-1022 (A change of counting method has no effect)
- Uddin, S. *et al.*, pp. 1166-1177 (A change of counting method has no effect)
- Wildgaard, L., pp. 1055-1078 (Included in the analysis)
- Yuret, T., pp. 1196-1206 (Included in the analysis)

**From *Scientometrics*, Volume 106, Issue 1 (January 2016)**

- Abramo, G. *et al.*, pp. 119-141 (Included in the analysis)
- Babić, D. *et al.*, pp. 405-434
- Benevenuto, F. *et al.*, pp. 469-474
- Bhardwaj, R. K., pp. 299-317
- Boukacem-Zeghmouri, C. *et al.*, pp. 263-280
- Hsiehchen, D. *et al.*, pp. 453-456
- Igami, M. *et al.*, pp. 383-403 (Included in the analysis)
- Klincewicz, K., pp. 319-345 (Included in the analysis)
- Lewison, G. *et al.*, pp. 105-117 (Included in the analysis)
- Ling, X. *et al.*, pp. 41-50 (A change of counting method has no effect)
- Mongeon, P. *et al.*, pp. 213-228

- Pritychenko, B., pp. 461-468 (Included in the analysis)
- Sun, Y. *et al.*, pp. 17-40
- Valentin, F. *et al.*, pp. 67-90
- van den Besselaar, P. *et al.*, pp. 143-162 (Included in the analysis)
- van Leeuwen, T. N. *et al.*, pp. 1-16
- Wang, L., pp. 435-452

**From *Scientometrics*, Volume 108, Issue 1 (July 2016)**
- Amat, C. B. *et al.*, pp. 41-56
- Andrei, T. *et al.*, pp. 1-20 (A change of counting method has no effect)
- Brizan, D. G. *et al.*, pp. 183-200
- Fell, C. B. *et al.*, pp. 113-141 (Included in the analysis)
- Höylä, T. *et al.*, pp. 263-288 (Included in the analysis)
- Moed, H. F., pp. 305-314
- Morichika, N. *et al.*, pp. 221-241 (Included in the analysis)
- Paul-Hus, A. *et al.*, pp. 167-182 (A change of counting method has no effect)
- Perovic, S. *et al.*, pp. 83-111
- Sanz-Casado, E. *et al.*, pp. 243-261 (Included in the analysis)
- Thelwall, M. 337-347 (A change of counting method has no effect)
- Winarko, B. *et al.*, pp. 289-304 (A change of counting method has no effect)
- Zuccala, A. *et al.*, pp. 465-484 (A change of counting method has no effect)

**From *Scientometrics*, Volume 109, Issue 3 (December 2016)**
- Abramo, G. *et al.*, pp. 1711-1724
- Abramo, G. *et al.*, pp. 1895-1909 (A change of counting method has no effect)
- Abramo, G. *et al.*, pp. 2053-2065 (Included in the analysis)
- Abramo, G. *et al.*, pp. 2093-2117 (Included in the analysis)
- Aman, V. pp. 2195-2216
- Ayaz, S. *et al.*, pp. 1511-1524
- Cánovas Izquierdo, J. L. *et al.*, pp. 1665-1693
- de Stefano, E. *et al.*, pp. 1579-1591
- Ebrahimy, S. *et al.*, pp. 1497-1510 (A change of counting method has no effect)
- Enger, S. G. *et al.*, pp. 1611-1638
- Fukugawa, N, pp. 2303-2327
- Glänzel, W. *et al.*, pp. 2165-2179 (A change of counting method has no effect)
- Guevara, M. R. *et al.*, pp. 1695-1709
- Huang, Y. *et al.*, pp. 1547-1559
- Kawaguchi, D. *et al.*, pp. 1435-1454 (Included in the analysis)
- Lepori, B. *et al.*, pp. 2279-2301
- Lewison, G. *et al.*, pp. 1877-1893 (Included in the analysis)
- Lin, A. J. *et al.*, pp. 1455-1476
- Lindahl, J. *et al.*, pp. 2241-2262
- Maity, B. K. *et al.*, pp. 2031-2048 (A change of counting method has no effect)

- Möller, T. *et al.*, pp. 2217-2239 (Included in the analysis)
- Mas-Bleda, A. *et al.*, pp. 2007-2030
- Pan, X. *et al.*, pp. 1593-1610
- Piro, F. N. *et al.*, pp. 2263-2278 (Included in the analysis)
- Pritychenko, B., pp. 2067-2076
- Salimi, N. *et al.*, pp. 1911-1938
- Snijder, R., pp. 1855-1875 (A change of counting method has no effect)
- Sun, Y. *et al.*, pp. 1965-1978 (A change of counting method has no effect)
- Tijssen, R. J. W. *et al.*, pp. 2181-2194
- Yan, Z. *et al.*, pp. 1815-1833 (A change of counting method has no effect)
- Yang, D.-H. *et al.*, pp. 1989-2005 (A change of counting method has no effect)
- Yu, L. *et al.*, pp. 1979-1987 (A change of counting method has no effect)
- Zanotto, S. R. *et al.*, pp. 1789-1814 (Included in the analysis)

The following studies do also meet criteria 2) arguing explicitly for the choices of counting methods:
**From *Research Evaluation*, Volume 25, Issue 1 (January 2016)**
- Neufeld, J., pp. 50-61 (Included in the analysis)
- Ramos, A. *et al.*, pp. 94-106 (Included in the analysis)

**From *Research Evaluation*, Volume 25, Issue 4 (October 2016)**
- Bloch *et al.*, pp. 371–382 (Included in the analysis)

**From *Aslib Journal of Information Management*, Volume 68, Issue 1 (2016)**
- Kumar, S., pp. 19-32 (Included in the analysis)

**From *Research Policy*, Volume 45, Issue 6 (July 2016)**
- Kahn, S. *et al.*, pp. 1304-1322 (Included in the analysis)

**From *Journal of the Association for Information Science and Technology*, Volume 67, Issue 12 (December 2016)**
- Frittelli *et al.*, pp. 3051-3063 (Included in the analysis)